\documentclass[twocolumn]{article}
\usepackage{amsmath,amssymb,braket}
\usepackage{times}
\usepackage[margin=2cm]{geometry}
\usepackage{microtype}
\begin{document}

\title{Reply to the Comment on ``Partial conservation of seniority
in semi-magic nuclei''}

\author{Chong Qi}

\maketitle
{\bf Abstract:}
    The present Reply addresses the Comment as posted on arXiv (arXiv:2606.04137 [nucl-th], June 2026).

\vspace{1em}

The Comment in Ref. \cite{Kai26} raises several objections to the treatment of Ref.~\cite{Neergard2022} in Sect.~4.2 of our review \cite{Qi2026}. While we accept some of the points as valid complementary clarifications of wording and logical presentation, none of them identify a scientific error that would mislead the reader or misrepresent the substance of Ref.~\cite{Neergard2022}. The corrections are at the level of emphasis and phrasing rather than physics.
As correctly stated in the review and as we explain below, we do not find it of sufficient scientific value to further emphasise the subspace decomposition or basis transformation approach unless and until a unique operator is identified that can directly generate the partially seniority-conserved states. Without such an operator, the transformation of the basis, however correct, remains of limited additional insight beyond what is already available from the symbolic shell-model framework of Ref.~\cite{QianQi2018}. 

We note that throughout this Reply we refer to the special $I=4$ and $I=6$ states with $v=4$ as the \textit{partially seniority-conserved states}, consistent with the terminology adopted in the review \cite{Qi2026}, while Neerg\aa rd refers to them as the Escuderos--Zamick multiplets.

To provide brief context for general readers, we recall that there are exactly 18 allowed states with different angular momentum  (and, optionally, seniority) values for a system of four identical particles in a single $j=9/2$ shell (see, Table 1 in the Review). These correspond directly to the 18 basis vectors with a total magnetic quantum number of $M=0$ in the so-called $M$-scheme representation. In practice, shell-model calculations can be performed either numerically or symbolically, utilizing either the $M$-scheme or the $jj$-coupled scheme. When working in the $jj$-coupled scheme, states with good total angular momentum are typically constructed using coefficients of fractional parentage (CFPs) or through direct angular momentum projection from the $M$-scheme basis. It is important to note that a standard two-body interaction conserves total angular momentum (rotational invariance), leading to strictly zero non-diagonal matrix elements between states with different spin values. However, it can change the seniority quantum number ($v$) by $\Delta v = 0, \pm 2,$ or $\pm 4$. Furthermore, while states with different total spins do not mix, the non-diagonal matrix elements between individual $M$-scheme basis vectors are generally non-zero. Both Sect.~4.2 of our review and the present Comment frame their analyses primarily within this $M$-scheme framework.

As proven analytically \cite{Qi2012}, the Hamiltonian cannot mix the partially seniority-conserved states with other states of the same spin and different seniority, ensuring that the corresponding non-diagonal matrix elements are exactly zero. Because an arbitrary change of basis will redistribute the Hamiltonian matrix elements, one can easily construct subspaces where specific non-diagonal elements vanish. Consequently, great care must be taken when analyzing the 'zeroness' (or invariance) of these couplings to distinguish between a mathematical redefinition of the basis and an underlying dynamical symmetry. 

There currently exists no known operator or transformation that uniquely projects out the partially seniority-conserved states. While Ref.~\cite{Neergard2022} represents an important effort in this direction, it is worth noting that numerous alternative transformations can be straightforwardly defined using the symbolic operations introduced in Ref.~\cite{QianQi2018}. For example:
\begin{itemize}
    \item Angular-momentum projection: Applying an angular-momentum projection operator isolates the partially seniority-conserved states into a three-dimensional subspace.
    \item Angular-momentum and seniority projection: Combining the angular-momentum projection with a seniority operator further restricts these states to a two-dimensional subspace.
    \item Hamiltonian diagonalization: By diagonalizing any seniority-conserving Hamiltonian matrix, one can completely isolate these states and derive their wave functions explicitly, whether in terms of the $M$-scheme basis or via single- and two-particle CFPs.
\end{itemize}
Therefore, what remains missing from the field is not the knowledge of the states themselves—which are already well-characterized and easily accessible—but rather the identification of a fundamental operator that transparently reveals their possible underlying symmetry properties.

\paragraph{On the invariance of $\Phi_4$ to the interaction of the particles.}
Concerning the second paragraph of the Comment, when a subspace is "invariant" under an operator (in this case, the two-body interaction Hamiltonian), it means that the two-body interaction does not mix the states inside the $\Phi_4$ subspace with any states outside of it. If one examines the Hamiltonian matrix, the off-diagonal elements connecting the states in $\Phi_4$ to the states in its orthogonal complement ($\Phi_4^\perp$) are exactly zero. The author states that proving this lack of mixing (the invariance of $\Phi_4$) is mathematically equivalent to proving the ``stationarity" (the unmixed, partially seniority-conserved nature).

This is unfortunately exactly the source of the misunderstanding. In general, as mentioned in the beginning, if one divides a set of states into two subgroups such that there is zero spin overlap between them (for example, Subgroup A has only spins $\{0, 2\}$ and Subgroup B has only spins $\{4, 6, 8\}$), those subgroups are trivially invariant under any rotationally invariant two-body interaction. What is exceptional and confusing about the specific partially seniority-conserved states is that they are immune to such divisions and can be placed in any above subgroups without affecting the invariance. In other words, one does not easily notice their presence just from the invariance of the two body interaction.

\paragraph{On the membership of the multiplets
in $\Phi_4^\perp$.}

Neerg\aa rd correctly notes in the 2nd and 3rd paragraph of the Comment that the partially seniority-conserved states belong to the $\Phi_4^\perp$ subspace as he defined it. We didn't dispute this fact or question the correctness of his operator; indeed, our review explicitly provides proper credit to the original paper for establishing this property.

\paragraph{On the basis vectors of $[\Phi_1]$ and their verification.}

Neerg\aa rd explains that the basis vectors of $\Phi_{40}^\perp$
were obtained by solving 14 homogeneous linear equations in 18
unknowns arising from orthogonality to $\Phi_{40}$, and that
diagonalising his matrix $C$ (Eq.~(22) of Ref.~\cite{Neergard2022})
in this basis reproduces the angular-momentum eigenstates listed
in our Table~4.
Because the transformed $M$-scheme vectors provided in Ref.~\cite{Neergard2022} do not carry definite angular momentum by construction, a direct comparison with our angular-momentum-projected basis inherently requires some additional diagonalisation steps. However, as the author's own description implies, performing these operations relies on a rather cumbersome and opaque procedure (essentially relying on manual maneuvers), which became apparent when we attempted to retrace them. While we do not question the mathematical correctness of this approach, its broader scientific significance and methodological utility should not be overemphasized.

With all due respect, we feel obliged to stand by this judgment, even if the author disagrees.

\paragraph{On the remark concerning unitary transformations.}

Neerg\aa rd points out that the transformation from the $M$-scheme
basis to the union of bases for $\Phi_{40}$ and $\Phi_{40}^\perp$ is
not unitary because the resulting basis is not orthonormal.
He further remarks that the caution we expressed is in any case
immaterial here because his subspaces $\Phi_4$, $\Phi_4^\perp$,
$\Phi_{40}$, and $\Phi_{40}^\perp$ are not direct sums of complete
angular-momentum eigenspaces. We maintain, however, that the caution expressed in our review is entirely warranted and physically sound. Any transformation of basis vectors must be handled with great care, particularly when the mathematical procedure used to generate it is physically obscure, as is the case in Ref.~\cite{Neergard2022}. Furthermore, one cannot simply bypass the necessity of a unitary transformation by adopting a non-orthonormal basis. In quantum mechanics, physical basis states must ultimately be orthonormal to one another to represent meaningful probability amplitudes—whether this orthonormalization is done explicitly during the construction of the basis or implicitly during the evaluation of observables. Otherwise, an arbitrary, non-orthonormal mathematical regrouping of the basis offers limited physical insight compared to a transparent symmetry operator.

\paragraph{On the omission of the main result.}

Neerg\aa rd states that what he considers the most remarkable result
of Ref.~\cite{Neergard2022} --- namely, that every rotationally
invariant two-body interaction acts on $\Phi_4^\perp$ as a linear
combination of a constant and the total angular-momentum operator
squared, $I^2$ --- was not mentioned in our review.
However, as we established in the beginning regarding invariant subspaces, this result is essentially a trivial mathematical consequence of the regrouping of states with different spin values (a step performed implicitly in Ref.~\cite{Neergard2022}). This algebraic reduction holds irrespective of the specific location of the partially seniority-conserved states. In other words, the exact same conclusion would emerge even if one were to arbitrarily move these specific states into the $\Phi_4$ group or separate them entirely. This behavior is an inherent byproduct of the subspace construction rather than a profound physical property unique to the partially seniority-conserved states.

\paragraph{Broader context: the significance of the subspace
decomposition.}
We feel obliged to offer
some broader context regarding the significance of the subspace
construction.
The decomposition of the full 18-dimensional $M = 0$ space into
$\Phi_{40}$ (dimension 14) and $\Phi_{40}^\perp$ (dimension 4)
is a natural and, we would argue, relatively straightforward
consequence of the angular-momentum and seniority structure of
the $(j=9/2)^4$ system.
In particular, such a decomposition emerges directly from the
symbolic shell-model approach of Ref.~\cite{QianQi2018}, which is
based on angular-momentum projection from $M$-scheme bases.
Starting from a trial wave function and projecting onto good
angular momentum, one automatically identifies the subspace
spanned by states with different spins and its complement, without the
need for a separate construction.
In this sense, while the subspace structure identified in
Ref.~\cite{Neergard2022} is correct, it does not provide new
computational leverage beyond what is already available in the
symbolic shell-model framework.

The central open challenge in this area remains the identification
of a \emph{unique operator} that projects out the partially
seniority-conserved $\alpha_1$ states directly --- analogous to the
pair creation operator $P_j^\dagger$ for the $v=0$ state, or the
angular-momentum projection operator $P_M^I$ for states with good
$I$.
Until such an operator is found, the subspace decomposition, however
formally correct, does not yield an operatorial understanding of why
these particular $v=4$, $I=4$ and $6$ states are singled out by
any two-body interaction.
This is the sense in which we described the problem as still open
in Sect.~4.2 of our review.

\end{document}